\documentclass[pdftex,twocolumn,epjc3]{svjour3}          

\usepackage[T1]{fontenc}

\smartqed  
\usepackage{amsmath,amssymb}
\usepackage{graphicx}
\usepackage{mathptmx}      
\usepackage{flushend}
\usepackage[numbers,sort&compress]{natbib}
\usepackage[colorlinks,citecolor=blue,urlcolor=blue,linkcolor=blue]{hyperref}
\usepackage{bm}

\begin{document}

\title{On Non-Perturbative Unitarity in Gravitational Scattering}

\author{Ivo Sachs$\,{}^{1}$
        \and
        Tung Tran$\,{}^{1,2}$
}

\thankstext{1}{\underline{e-mail}: ivo.sachs@physik.lmu.de}
\thankstext{2}{\underline{e-mail}: tung.tran@physik.uni-muenchen.de; tung.tran@aei.mpg.de}

\institute{${}^{1}$Arnold Sommerfeld Center for Theoretical Physics,\\
 Ludwig Maximilian University of Munich,\\
Theresienstr.\enspace 37, D-80333 M\"unchen,\enspace Germany\\~\\
${}^{2}$Albert Einstein Institute,\\
Am M\"uhlenberg 1, D-14476, Potsdam-Golm, Germany
}

\date{}

\maketitle
\begin{abstract}
We argue that the tree-level graviton-scalar scattering in the Regge limit is unitarized by non-perturbative effects within General Relativity alone, that is  without resorting to any extension thereof. At Planckian energy the back reaction of the incoming graviton on the background geometry produces a non-perturbative plane wave which softens the UV-behavior in turn. Our amplitude interpolates between the perturbative graviton-scalar scattering at low energy  and  scattering on a classical plane wave in the Regge limit that is bounded for all values of $s$.
\end{abstract}
\section{\label{Intro}Introduction}
It is well known that perturbative scattering amplitudes involving gravitons violate the unitarity bound at Planckian energy even at tree-level. For instance, the scattering amplitude of a graviton and a massless scalar field is given by \cite{Berends:1974gk}
\begin{align}\label{eq:berendsamplitude}
    A_{tree}=-(\epsilon_1\cdot\epsilon_3)\frac{i\kappa^2}{4}\frac{su}{t}\,,
\end{align}
where $s,t,u$ are the usual Mandelstam variables and $\kappa$ is the dimensionful gravitational coupling, grows without bound as $s$ increases at fixed $t$. This state of affairs has given rise to an extensive activity in searching for a UV-completion of General Relativity (GR). String theory is one such complete theory whose legacy rests partly on the fact that it predicts an amplitude that is perturbatively unitary.   

\begin{figure}[h]
    \centering
    \includegraphics[scale=0.33]{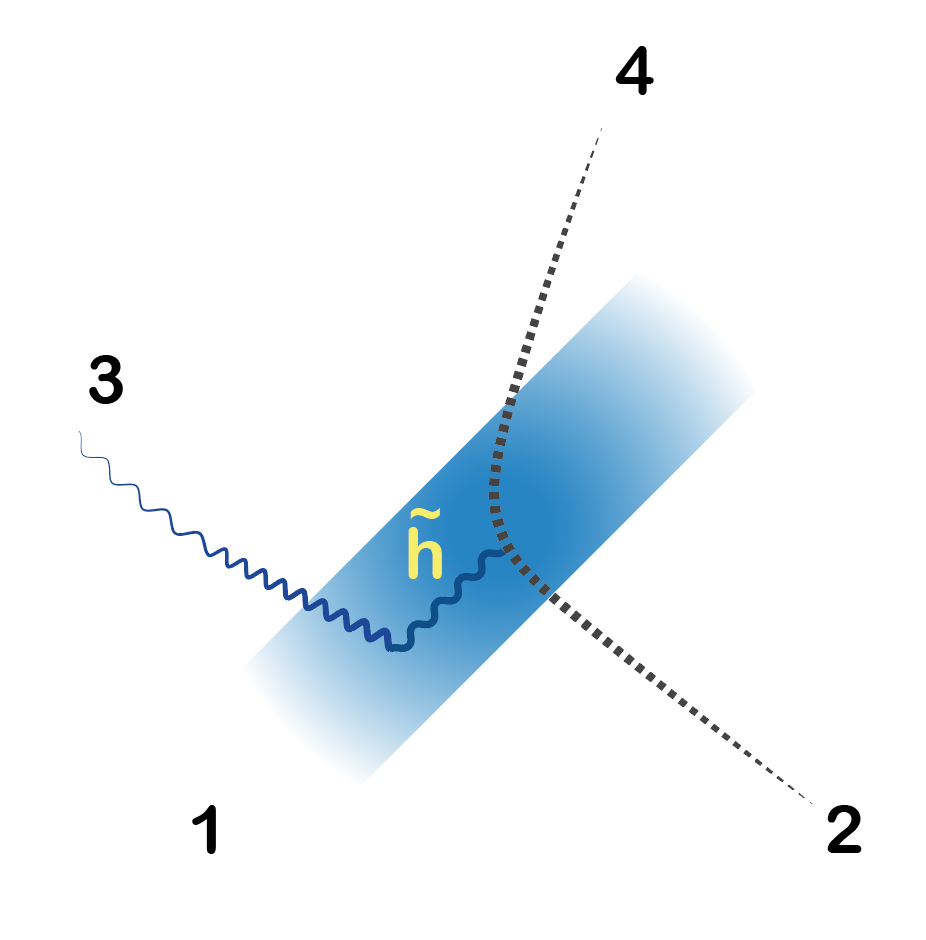}
    \caption{Non-perturbative four particle  scattering process where the ingoing gravitons is replaced by a plane wave (shaded strip).}
    \label{fig:4ptonpp}
\end{figure}

On the other hand, one may question whether the assumption of asymptotic in- and out-states on which \eqref{eq:berendsamplitude} is built holds for gravitons of Planckian energy since Gravity is a non-linear theory whose coupling strength increases with energy. One argument in favor of it is that a single graviton can always be boosted to an inertial frame where its energy is small. However, for a two body scattering with large center of mass (CoM) energy $\sqrt{s}$, there is no boost for which both particles have small energy. Thus back-reaction will have to be taken into account for at least one in-going particle. This idea is not new. It was explored already many years ago by 't Hooft \cite{tHooft:1987vrq} and others \cite{Ferrari:1988cc} who replaced an ingoing scalar of transplanckian energy by a gravitational shock wave. One may also interpret this back-reaction as a contribution to the self completeness mechanism of gravity proposed by Dvali and Gomez \cite{Dvali:2010bf,Dvali:2011th}. The starting point on which we base our argument for a non-perturbative unitarization of \eqref{eq:berendsamplitude} is similar to \cite{tHooft:1987vrq} although the details are somewhat different. We perform a Lorentz boost such that the energy of the incomming scalar is small while the incomming graviton has transplanckian energy so that back reaction on geometry has to be taken into account. Luckily, an exact solution to Einstein's equation, accounting for the complete back reaction on geometry is available in the form of a plane wave \cite{,Brinkmann:1925pp,Einstein:1937qu} (c.f. \cite{Penrose:1965pw,Garriga:1990dp,JGrif:1991JG,MBlau:2011mb,Stephani:2003tm}). As a result, the non-perturbative generalization of \eqref{eq:berendsamplitude} in the large $s$ but small $t$ (or Regge) limit can be reduced to a perturbative calculation on top of a plane wave as illustrated in Fig. \ref{fig:4ptonpp}. 

\section{\label{secPT}Perturbative limit}
To see how this comes about let us first recover the perturbative amplitude \eqref{eq:berendsamplitude} for $( h_{(1)}, \phi_{(2)}\to h_{(3)}, \phi_{(4)})$ in position space. 
Without restricting the generality we make the following momentum assignments  
\begin{align}
p_{(1)}\equiv &p=(p_+,0,0,0)\;,\quad \quad \,
p_{(2)}\equiv k=(0,k_-,0,0)\nonumber\\
p_{(3)}\equiv &q=(q_+,q_-,q_1,q_2)\;,\quad 
p_{(4)}\equiv l\,=(l_+,l_-,l_1,l_2)\,.\nonumber
\end{align}
In position space the $t$-channel diagram can then be calculated as follows: We first solve for the internal graviton $\tilde h$ around Minkowski background, $\eta$, through
\begin{align}\label{ht}
0&=G_{\mu\nu}(\eta+\lambda h_{(1)}+\lambda h_{(3)}+\lambda^2 \tilde h)\nonumber\\
&=\lambda^2\frac{\delta G_{\mu\nu}}{\delta g}\Big|_\eta  (\tilde h) + \lambda^2\frac{\delta^2 G_{\mu\nu}}{\delta g^2}\Big|_\eta (h_{(1)},h_{(3)})\,,
\end{align}
where $G_{\mu\nu}$ is the Einstein tensor and we assume that $h_{(1)}$ and $h_{(3)}$ satisfy the linearized Einstein equation 
\begin{align}
    \frac{\delta G_{\mu\nu}}{\delta g}\Big|_\eta  (h_{(1)})=  \frac{\delta G_{\mu\nu}}{\delta g}\Big|_\eta  (h_{(3)})&=0\,.
\end{align}
 Here, $\lambda$ is a dimensionless parameter whose sole purpose is to keep track of the order in perturbation in $h$. Next, we solve for the outgoing scalar field $\tilde\phi$ with the help of the Ansatz $\phi=\phi_{(2)}+\lambda^2 \tilde\phi$, 
\begin{eqnarray}\label{pt}
0=\Box_{\eta+\lambda^2 \tilde h}\phi=\lambda^2 \Box_{\eta} \tilde\phi+\Box_{\lambda^2\tilde h}\phi_{(2)}\,,
\end{eqnarray}
where $\Box_{g}$ stands for scalar wave operator in the metric background $g$. We note that (\ref{ht}) fixes $\tilde h$ only up to a solution of the homogeneous equation. The latter  reproduces 3-particle (1 graviton) scattering amplitude upon substitution into \eqref{pt}. Note also that in (\ref{ht}) we can replace $ h_{(1)}$ by a wave packet since the equation for $\tilde h$ is linear in $ h_{(1)}$.

Equivalently, we can treat $h_{(1)}$ as a background field and solve $\tilde h$ as a linearized fluctuation around that background. Setting $h_{(1)}\equiv H$ for later convenience, the linearized Einstein equation reads
\begin{eqnarray}\label{hht2}
0=\frac{\delta G_{\mu\nu}}{\delta g}\Big|_{\eta+\lambda  H}(h) \,.
\end{eqnarray}
With the Ansatz \footnote{The choice of the linear contribution is fixed by the initial condition to have an ingoing graviton $h=\lambda h_{(3)}$.} $h=\lambda h_{(3)} +\lambda^2 \tilde h$, we expand the background once again and get
\begin{eqnarray}\label{ht2}
0=\lambda^2\frac{\delta G_{\mu\nu}}{\delta g}\Big|_\eta  (\tilde h) +\lambda^2\frac{\delta^2 G_{\mu\nu}}{\delta g^2}\Big|_\eta (H,h_{(3)}),
\end{eqnarray}
which is in agreement with (\ref{ht}). 

In what follows we will work with the latter form since it is suitable to accommodate non-linear effects for the incomming graviton, $H$. Indeed, suppose that $ H$ has momentum $p$ of order $M_{Pl}$. Then  $H$ cannot be treated as a perturbation of Minkowski space-time and back reaction on the geometry has to be taken into account. This can be done by replacing $ H$ by a plane wave. In Einstein-Rosen coordinates \cite{Einstein:1937qu} the plane wave metric reads 
\begin{align}
    ds^2=2dy^+dy^--\gamma_{ij}(y^+)dy^idy^j, \quad (i,j=1,2),
\end{align}
with $\gamma_{ij}\sim (\delta_{ij}+ h_{(1)ij})$ in the perturbative limit (here and in what follows we absorb $\lambda$ in $h$). However, in the non-perturbative regime, Brinkmann coordinates \cite{Brinkmann:1925pp} are more convenient, with
\begin{equation}\label{eq:Brinkmann}
    ds^2=2dx^+dx^- - H_{ab}(x^+)x^ax^b(dx^+)^2 -dx_a^2, \quad (a,b=1,2)
\end{equation}
which is an exact solution, $G_{\mu\nu}(\eta+H)=0$. Then  (\ref{hht2}) is the correct generalization of  (\ref{ht}) provided $h_{(3)}$ has small momentum which is compatible with the Regge limit, $t\ll M_{Pl}^2$. In Brinkmann coordinates, for  $h_{(3)}$ transverse, traceless with asymptotic polarization vector $\epsilon_a$, the linearized solution for $\tilde h$ on the plane wave takes the form \cite{Adamo:2017nia}
\begin{eqnarray}\label{thp}
\tilde h_{\mu\nu}=\begin{pmatrix}0&0&0&0\cr0&-\frac{i}{q_-}\epsilon^a\Sigma_{ab}\epsilon^b +\epsilon_+^2& \epsilon_+ \epsilon_1&\epsilon_+ \epsilon_2\cr 0& \epsilon_+\epsilon_1&\epsilon_1\epsilon_1&\epsilon_1\epsilon_2\cr
0&\epsilon_+\epsilon_2&\epsilon_2\epsilon_1&\epsilon_2\epsilon_2\end{pmatrix}\Phi(x)
\end{eqnarray}
where we have chosen the light-cone gauge ($h_{-\mu}=0$),
\begin{eqnarray}\label{eq:ppscalar}
\Phi(x)=\frac{1}{\sqrt{|E|}}\, e^{iq_-\big(x^-+\frac{\Sigma_{ab}x^ax^b}{2}\big)+iq_iE^i_ax^a+i\frac{q_iq_j}{q_-}F^{ij}(x^+)}
\end{eqnarray}
is the solution of the scalar wave equation and 
\begin{equation}\label{F}
    \Sigma_{ab}=\dot{E}^i_{\ a}E_{bi}, \quad F^{ij}(x^+)=\int_{-\infty}^{x^+}\gamma^{ij}(\tau)d\tau.
\end{equation}
Here $\Sigma_{ab}$ is the deformation tensor for the Vierbein ${E}^i_{\ a}$, subject to 
\begin{equation}\label{eq:ppwaveH}
    \ddot{E}_{ai}=H_{ab}(x^+)E^{b}_i\,, \quad\text{with}\quad\lim_{x^+\rightarrow \infty}E^i_a(x^+)=\delta^i_a\,,
\end{equation}
and $\gamma_{ij}={E}_{i a}E^a_{\ j}$. Finally, the longitudinal polarization $\epsilon_+$ is given by
\begin{equation}\label{eu}
\epsilon_+=\epsilon^a\Big[\frac{q_j}{q_-}E_a^j+\Sigma_{ab}x^b\Big]
\,.
\end{equation}
In order to disentangle the disconnected 1-graviton contribution, we then substitute \eqref{thp} into \eqref{pt}
\begin{align}\label{bp1}
\Box_{\eta}\tilde\phi&=\partial_-\Big[\epsilon_+^2-\frac{i}{q_-}\epsilon^a \epsilon^b\Sigma_{ab}\Big]\Phi \partial_-\phi_{(2)}\\
&-\partial_-\epsilon_+ \epsilon^a\Phi\partial_a\phi_{(2)}-\partial_a\epsilon_+ \epsilon^a\Phi\partial_-\phi_{(2)}+\partial_a\epsilon^a\epsilon^b\Phi\partial_b\phi_{(2)}\nonumber
\end{align}
with $\phi_{(2)}$ is the incoming scalar field. We can then make connection to perturbation theory around Minkowski metric by integrating $\Box_{\eta}\tilde\phi$ against $\phi_{(4)}$,
\begin{align} \label{upm}
\int\limits_{R^4}\phi_{(4)}\Box_{\eta}\tilde\phi&=\lim_{x^+\to\infty}\int\limits_{\Sigma_{x^+}}\left(\phi_{(4)}\partial_-\tilde\phi-\tilde\phi\partial_-\phi_{(4)}\right)\\
&\quad - \lim_{x^+\to-\infty}\int\limits_{\Sigma_{x^+}}\left(\phi_{(4)}\partial_-\phi_{(2)}-\phi_{(2)}\partial_-\phi_{(4)}\right)\nonumber
\end{align}
where we have used that $\tilde\phi\to \phi_{(2)}$ as  $x^+\to-\infty$. The second term on the r.h.s. then subtracts the disconnected contribution to the scattering. At $0^{th}$ order in $H $ this gives 
\begin{eqnarray}\label{3s}
-(2\pi)^4\delta^4(k+q+l)(\epsilon^\mu k_\mu)^2\,,
\end{eqnarray}
thus reproducing the familiar $1$-graviton scattering, as expected, since $\tilde h$ reduces to $h_{(3)}$ in (\ref{ht}) for vanishing $H$.

Let us now consider the first order in $H$. We first choose a polarization for $H$ by setting 
\begin{equation}
E^i_a=\delta^i_a+\hat\sigma^i_a\;e^{ip_\mu x^{\mu}},
\end{equation}
where $\hat\sigma^i_a$ is one of the Pauli matrices. The contributions, linear in $H$, come from the expansion of $\epsilon_+$ and $\Sigma_{ab}$ in (\ref{bp1}). To continue we note that the deformation in $\epsilon_+$ simply takes account of the fact that the transversality condition of $h_{(3)}$ depends on $H$, so that the $\epsilon_+$ contribution is most naturally interpreted as a deformation of the 3-pt amplitude \eqref{3s}. The contribution at first order in  $H$ to the connected four point function is then
\begin{equation}
    -(\epsilon^a\epsilon^b\hat\sigma_{ab})\;p_+\frac{k_-^2}{q_-}
\end{equation}
where we used that $\Sigma_{ab}=ip_+ \hat\sigma_{ab}$. In terms of the Mandelstam variables this can be written as 
\begin{equation}\label{pt111}
    -(\epsilon^a\epsilon^b\hat\sigma_{ab})\frac{s^2}{t}
\end{equation}
which is the $t$-channel contribution of the scalar-graviton into scalar-graviton scattering amplitude. 

Next we replace $\phi_{(2)}$ by the linearized approximation of the scalar solution $\Phi_{(2)}$ in the  plane wave background,
\begin{eqnarray}\label{Phisim}
\Phi_{(2)}(x)=\frac{e^{ik_-\big[x^-+\frac{\Sigma_{ab}x^ax^b}{2}\big]}}{\sqrt{|E|}}\sim\left[1+ \frac{i\Sigma_{ab}x^ax^b}{2}\right]e^{ik_-x^-}.\nonumber
\end{eqnarray}
This should, in addition, account for the the $s$ and $u$- channel contribution. Indeed, solving for the scalar wave equation in the plane wave background before and after interacting with $ \tilde h$ takes into account the interaction with $H$. Furthermore, this should account for the contact interaction. Indeed, the last term in (\ref{bp1}) gives an extra contribution 
\begin{equation}
    -(\epsilon^a\epsilon^b\hat\sigma_{ab})\;p_+k_-= -(\epsilon^a\epsilon^b\hat\sigma_{ab})\;s\,.
\end{equation}
Adding this to (\ref{pt111}) we get 
\begin{equation}\label{pt11}
 (2\pi)^4\delta^4(p+k+q+l)   (\epsilon^a\epsilon^b\hat\sigma_{ab})\frac{su}{t}
\end{equation}
which is the correct perturbative limit including all channels as well as the contact interaction. 

\section{Non-perturbative calculation}
In order to take the complete backreaction of the incomming graviton into account we make the substitution $\eta\to \eta+H$ and insert the exact solutions $\Phi_{(2)}$ and $\Phi_{(4)}$ for the scalar fields together with the internal graviton $\tilde h_{\mu\nu}$ on the plane wave into \eqref{bp1}. Then integrating \eqref{bp1} against $\Phi_{(4)}$ we end up with (ignoring the one graviton contribution (\ref{3s})). 
\begin{align}
A&=\int \Phi_{(4)} \partial_\mu \tilde h^{\mu\nu}\partial_\nu\Phi_{(2)}\nonumber\\
&=\int \Phi_{(4)}\Big(\frac{i}{q_-}\epsilon^a\epsilon^b\Sigma_{ab}k_-^2+ik_-\epsilon^a\epsilon^b\Sigma_{ab}\Big)\Phi\,\Phi_{(2)},
\end{align}
where the terms in the bracket come form the first line in \eqref{bp1} and the integral at $x^+\to-\infty$ in (\ref{upm}) subtracts the disconnected contribution, with $h_{(3)}$ now disconnected in the plane wave background. This integral can be further simplified following \cite{Adamo:2017nia},
\begin{align}
A&=-(2\pi)^2\delta(k_-+q_-+l_-)\int du\frac{(\epsilon^a\epsilon^b\Sigma_{ab})\left(\frac{k_-^2}{q_-}+k_-\right)}{|E||E_{(2)}|^{\frac{1}{2}}\sqrt{|l_- B|}}\\
&\qquad\times \exp\Big[{i\Big(-\frac{J_a(B^{-1})^{ab}J_b}{2l_-}+\frac{q_iq_j}{q_-}F^{ij}+\frac{l_il_j}{l_-}F^{ij}\Big)}\Big]\,,\nonumber
\end{align}
where $E_{(2)}$ again solves \eqref{eq:ppwaveH} but with {\it ingoing} boundary condition, $\lim\limits_{x^+\rightarrow -\infty}(E_{(2)})^i_a(x^+)=\delta^i_a$, and 
\begin{align}
    J_a=(l_i+q_i)E^{i}_a, \quad B_{ab}=\Sigma_{ab}^{(2)}-\Sigma_{ab}.
\end{align}
To continue we note that $E^i_a$ are functions of $z=p_+ x^+$ only while $\Sigma_{ab}$, and $B$ are of the form $p_+ \boldsymbol{\Sigma}(z)$, $p_+ \boldsymbol{B}(z)$ respectively 
. This allows us to extract the $p_+$ dependence as 
\begin{align}\label{uni1}
A&=(2\pi)^2\delta(s+u+t)\frac{s}{t}\int dz\frac{(\epsilon^a\epsilon^b\boldsymbol{\Sigma}_{ab})}{|E||E_{(2)}|^{\frac{1}{2}}}\frac{1}{\sqrt{| \boldsymbol{B}|}}\\
&\ \ \ \times \exp\Big[{i\Big(-\frac{J_a(\boldsymbol{B}^{-1})^{ab}J_b}{u}+2\big(\frac{l_il_j}{u}+\frac{q_iq_j}{t}\big)F^{ij}}\Big)\Big]\,,\nonumber
\end{align}
where the extra factor of $1/p_+$ multiplying $F^{ij}$ in the last term is due to the change of measure $d\tau=\frac{dz}{p_+}$ in \eqref{F}. In addition we used that 
\begin{align}\label{uswp}
    \delta(s+u+t)=\delta((k_-+q_-+l_-)\cdot p_+)=\frac{\delta(k_-+q_-+l_-) }{p_+}
\end{align}
where $2q_-p_+=t,\;2k_-p_+=s$. This relation shows that the prefactor $\delta(s+u+t)\frac{s}{t}$ in (\ref{uni1}) is bounded for transplanckian values of $p_+$ and therefore also in $s$. Using (\ref{uswp}), the form (\ref{uni1}) of the 4-point amplitude makes the unitarity of the amplitude at large $s$ manifest. Indeed, the integral in (\ref{uni1}) is absolutely convergent for any value of $u$ (as we will see below). On the other hand for small $s$ \eqref{uni1} reduces to the perturbative amplitude (\ref{pt11}). Thus the amplitude \eqref{uni1} is the non-perturbative, unitary completion of \eqref{eq:berendsamplitude}.

It is not hard to see that the terms in \eqref{thp} containing $\epsilon_+$ and $\epsilon_a$  will similarly give a non-perturbative deformation of \eqref{3s} preserving the unitarity of the latter.  

We would like to stress, however, that the boundedness of the scattering amplitude does not imply that the total cross-section for the $(\phi(k),h_{\mu\nu}(p)\to \; \phi(l),h_{\mu\nu}(q))$ scattering is unitary since, due to the absence of momentum conservation, the integral over the outgoing momenta is not constrained. However, this feature is expected for scattering on an external potential. What our calculation shows then is that the question of unitarity of the gravitational four point scattering is actually not well posed. What we find is that at large center of mass energy, back reaction builds up an external field (the plane wave) so that at large $s$ and small $t$, the four-point scattering is actually better described by a scattering off an external plane wave.

In order to complete the argument that the amplitude is bounded we need to convince ourselves that the integral in (\ref{uni1}) is finite. As mentioned before, all the steps performed in obtaining (\ref{uni1}) are equally valid when replacing $h_{(1)}$ by a wave packet which is the more realistic set-up. Let us then consider the particular case when the plane wave is a sandwich wave \cite{Penrose:1965pw}, that is, it vanishes for $|x^+|>x_0$ where $x_0>0$. A generic feature of such plane waves is the focusing of geodesics \cite{Penrose:1965pw,Garriga:1990dp} which implies, in particular, that $|\gamma_{ij}|$ vanishes at some point $x^+>x_0$. Consequently 
the amplitude of (\ref{eq:ppscalar}) will be singular at this point and so will $\boldsymbol{\Sigma}_{ab}(z)$ and $\boldsymbol{B}(z)$. We can further simplify to the case where the plane wave is delta function supported in $u$ with linear polarization. In this case we have 
\begin{eqnarray}\label{deltaH}
H_{ab}(x^+)\propto \delta(x^+)\begin{pmatrix}1&0\cr 0&-1\end{pmatrix}\,.
\end{eqnarray}
It is a simple matter to show (e.g. \cite{Garriga:1990dp}) that $E^a_2$ has a simple zero (and thus $\gamma_{22}$ has a double zero) while $E^a_1>~0$. Therefore, the zero  of $|E_{(2)}(z)|$ and the pole $\boldsymbol{B}(z)$ cancel against each other so that we are left with a simple pole coming from $\boldsymbol{\Sigma}_{ab}(z)$. Thus, the integral exists in the sense of distributions and is bounded in the CoM energy, $s$ (and also in $u$).

The remaining terms in which lead to the 3-point amplitude (\ref{3s}) in the perturbative limit will also receive non-perturbative contributions upon replacing $\phi_{(2)}$ and $\phi_{(4)}$ by the exact solution in the plane wave background. It is not hard to see that that these are bounded in the large $s$ limit. 

\section{Back Reaction}\label{backreaction}
So far we have ignored the backreaction of the scattered particles on the plane wave. On the other hand, due to the focusing of the geodesics we expect that the energy-momentum density of the matter (and gravitons) will diverge at the focusing points \cite{Garriga:1990dp}. That this is indeed the case can be seen by recalling that the scalar field $\Phi_{(2)}(x)$ in the plane wave background given in (\ref{eq:ppscalar}). From this we see that the amplitude of $\Phi_{(2)}(x)$ grows like $|\gamma_{ij}|^{-\frac{1}{4}}\sim \frac{1}{\sqrt{x^+}}$ near the focusing point which we take to be located at $x^+=0$ for convenience. The dominant contribution to the stress tensor thus comes from 
\begin{eqnarray}
T_{++}\sim \partial_+\Phi_{(2)}\partial_+\Phi_{(2)}\sim \frac{1}{(x^+)^3}\,.
\end{eqnarray}
Thus back reaction is important near the focusing point and (\ref{deltaH}) should be modified accordingly. In order to obtain a self-consistent solution let us define $\mathcal{E}:=|E|=\sqrt{|\gamma_{ij}|}$. Then we have \cite{Garriga:1990dp} 
\begin{eqnarray}
\frac{{\ddot{\mathcal{E}}}}{\mathcal{E}}=-\frac{R_{++}}{4}-\frac{\text{tr}((\gamma^{-1}\dot{\gamma})^2)}{16}+\frac{\text{tr}((\gamma^{-1}\dot{\gamma}))^2}{32}-\frac{\dot{\mathcal{E}}^2}{\mathcal{E}^2}
\end{eqnarray}
with 
\begin{equation}\label{AST}
    R_{++}=T_{++}\sim(\partial_+\frac{1}{\sqrt{\mathcal{E}}})^2\sim \frac{1}{4}{\frac{\dot{\mathcal{E}}^2}{\mathcal{E}^3}}\,.
\end{equation}
Near the focusing point, where ${\mathcal{E}}$ vanishes, the curvature term dominates so that near $x^+=0$, by rescaling $\mathcal{E}\rightarrow\frac{\mathcal{E}}{16}$, we get 
\begin{eqnarray}
\frac{{\ddot{\mathcal{E}}}}{\mathcal{E}}= -{\frac{\dot{\mathcal{E}}^2}{\mathcal{E}^3}}
\end{eqnarray}
which has a first integral, $\dot{\mathcal{E}}=e^{\frac{1}{{\mathcal{E}}}}$. 
Let us then consider $z$ as a function of ${\mathcal{E}}$, that is, \begin{eqnarray}\label{ieq}
\frac{dz}{d\mathcal{E}}=e^{-\frac{1}{{\mathcal{E}}}}\,.
\end{eqnarray}
Upon substitution into (\ref{uni1}) focusing on the pre-exponentional factor near the zero-locus of ${\mathcal{E}}$, as before, we find 
\begin{eqnarray}
\int dz (\cdots) \sim \int\frac{\sqrt{\dot{\mathcal{E}}}}{{\mathcal{E}}}dz= \int \frac{e^{-\frac{1}{2{\mathcal{E}}}}}{{\mathcal{E}}}d{\mathcal{E}}
\end{eqnarray}
which is finite near the focusing point.

Before we close this section we should comment on the justification of \eqref{AST} which we claimed to be the dominating term near the focusing point. This is apparent when expressing $\Phi_{(2)}$ in Einstein-Rosen coordinates
\begin{eqnarray}
\Phi_{(2)}(x)=\frac{e^{ik_-y^-}}{\sqrt{|E|}}.\nonumber
\end{eqnarray} 
On the other hand, the actual calculations are done in Brink\-mann coordinates where the phase of the scalar field \eqref{Phisim} oscillates rapidly near $x^+=0$. So one might argue that a more singular contribution to $T_{++}$ comes form differentiating the phase. However, this is clearly an artifact of the choice of Brinkman coordinates. Indeed the coordinate transformation 
\begin{align}\label{eq:diffeo}
    y^+&=x^+,\nonumber \\
    y^-&=x^-+\frac{1}{2}\dot{E}^i_{\ a}E_{bi}x^ax^b=x^-+\frac{1}{2}\Sigma_{ab}x^ax^b, \\
    y^i&=E^i_ax^a,\nonumber
\end{align}
is singular at the focusing point away from $x=0$. In Rosen coordinates this rapid oscillation is simply expressed by noticing that for $x\neq 0$, $x^+\to 0$ maps into $y^-\sim \frac{x^2}{x^+}\to\infty$ which is not a singular point. If we now consider a wave packet of compact support in the $y^-$-direction which will cut-off the wave function, at fixed $x^-$ in the $x$-direction in agreement with causality. On the other hand the pre-exponential factor accounting for \eqref{AST} is not a coordinate artifact. It simply reflects the focusing of the geodesics which is a geometric property of plane waves \cite{Penrose:1965pw}.
\section{Discussion}
We have shown that the perturbative 4-particle amplitude evolves in the large $s$, small $t$ limit into a non-perturbative expression involving a macroscopic non-linear plane wave, which is manifestly unitarity at the expense of smearing out the momentum conservation constraint. This picture is intuitively satisfactory, since due to backreaction, we expect that an energetic graviton sources a growing number of soft gravitons, eventually approaching a classical solution. Earlier approaches based on related ideas were proposed by 't Hooft and others  \cite{tHooft:1987vrq,Ferrari:1988cc} used a gravitational shock wave to represent an energetic scalar particle (in the geometrical optics approximation) and studied the propagation of a scalar field in that background. One might suggest that this setting should be related to ours via a boosted reference frame in which the incoming scalar $\phi_{(2)}$ is energetic while the graviton $h_{(1)}$ is perturbative. However, since the shock wave approximation \cite{tHooft:1987vrq} works only for point particle sources or superposition thereof \cite{Ferrari:1988cc}, the matching is not clear. In fact the scenario of \cite{tHooft:1987vrq} does not have simple perturbative limit. On the other hand, at the calculational level some of our formulas are essentially identical to those in \cite{Adamo:2017nia} but the physical interpretation is quite different. 

Finally, we should emphasize our result does not yet allow us to conclude that GR unitarizes itself completely since large momentum transfer, where new physics usually arises, is not covered by our analysis. Other approaches which focus on the large $t$ limit instead can be found in \cite{Dvali:2010bf} for instance, where it is argued that black holes may unitarize the cross section in the large $t$ limit. We have nothing new to say about that regime apart, perhaps, that in order to set up an \, experiment involving gravitons with large momentum transfer at least one of the ingoing gravitons must have energy of the order of $M_{Pl}$ in which case our analysis becomes relevant. The same comment applies, of course, to scattering at transplanckian energies in string theory \cite{Dvali:2014ila}. 

\section*{Acknowledgments}
The authors would like to thank C. Gomez as well as Tomas Prochazka for discussions and, in particular, S. Mukhanov for substantial input. This work has received support from the Excellence Cluster 'Origins: From the Origin of the Universe to the First Building Blocks of Life'.

\bibliographystyle{apsrev4-1}
\bibliography{pp.bib}
\end{document}